\documentstyle[11pt,epsf]{article}

\def\indspace{\hspace*{1.0em} }
\def\appendix{\setcounter{section}{0}
\def\thesection{Appendix \Alph{section}}
\def\theequation{\Alph{section}.\arabic{equation}}}

\newfont{\subsub}{cmr6}

\newcounter{szk}

\begin{document}
\title{Annual change of Pareto index dynamically deduced from\\ the law of detailed quasi-balance
}
\author{
\footnote{e-mail address: ishikawa@kanazawa-gu.ac.jp} Atushi Ishikawa
\\
Kanazawa Gakuin University, Kanazawa 920-1392, Japan
}
\date{}
\maketitle

\begin{abstract}
\indent
Employing data on the assessed value of land
in 1983 -- 2005 Japan,
we investigate the dynamical behavior in the high scale region
of non-equilibrium systems.
From the detailed quasi-balance
and Gibrat's law, we derive
a relation between the change of Pareto index and a symmetry in
the detailed quasi-balance.
The relation is confirmed in the empirical data nicely.
\end{abstract}
\begin{flushleft}
\end{flushleft}

\vspace{1cm}
\section{Introduction}
\label{sec-Introduction}
\indspace
In physics, 
universal laws in nature are extracted 
by analyzing empirical data,
and various phenomena are explained by the laws.
In econophysics, 
economic phenomena are investigated
by using this physical approach \cite{MS}.
From this point of view, recently
a major progress has been made with the research in distributions of personal income
or company size. 

More than a hundred years ago, it is well known that
a cumulative number $N(> x)$ obeys a power-law
for $x$ which is larger than a certain threshold $x_0$:
\begin{eqnarray}
    N(> x) \propto x^{-\mu}~~~~{\rm for }~~~~x > x_0~.
    \label{Pareto}
\end{eqnarray}
This power-law and the exponent $\mu$ are called 
Pareto's law and Pareto index, respectively \cite{Pareto} \footnote{
Pareto's law is checked with high accuracy (See \cite{ASNOTT, Yakovenko} for instance).
}.
Here $x$ is wealth, income, 
profits, assets, sales, the number of employees and etc.

Fujiwara et al. \cite{FSAKA} have explained this phenomenon
by using the law of detailed balance and Gibrat's law \cite{Gibrat},
which are observed in empirical data.
The detailed balance is time-reversal symmetry:
\begin{eqnarray}
    P_{1 2}(x_1, x_2) = P_{1 2}(x_2, x_1)~.
    \label{Detailed balance}
\end{eqnarray}
Here $x_1$ and $x_2$ are  two successive incomes, 
profits, assets, sales, etc, and
$P_{1 2}(x_1, x_2)$ is a joint probability distribution function (pdf).
Gibrat's law states that
the conditional probability distribution of growth rate $Q(R|x_1)$ 
is independent of the initial value $x_1$: 
\begin{eqnarray}
    Q(R|x_1) = Q(R)~.
    \label{Gibrat}
\end{eqnarray}
Here growth rate $R$ is defined as the ratio $R = x_2/x_1$ and
$Q(R|x_1)$ is defined by using the pdf $P_1(x_1)$
and the joint pdf $P_{1 R}(x_1, R)$ as
\begin{eqnarray}
    Q(R|x_1) = \frac{P_{1 R}(x_1, R)}{P_1(x_1)}~.
    \label{conditional}
\end{eqnarray}
In the proof, they assume no model and only use
these two underlying laws in empirical data.

Below the threshold $x_0$,
it is also well known that the power-law
is not observed \cite{Gibrat, Badger}.
\footnote{
It is pointed out that the reason is
the breakdown of Gibrat's law \cite{FSAKA}.
The breakdown of Gibrat's law in empirical data
is reported by Stanley's group \cite{Stanley1}.
Takayasu et al. \cite{TTOMS} and Aoyama et al. \cite{Aoyama} also report that
Gibrat's law does not hold in the middle scale region by using data of Japanese companies.
}
In Ref.~\cite{Ishikawa3}, 
by employing profits data of Japanese companies in 2002 and 2003 \cite{TSR},
Gibrat's law (\ref{Gibrat}) is extended in the middle scale region
as follows:
\begin{eqnarray}
    Q(R|x_1)&=&d~R^{-t_{+}(x_0)-\alpha_{+} \ln \frac{x_1}{x_0}-1}~~~~~{\rm for}~~R > 1~,
    \label{Rline1}\\
    Q(R|x_1)&=&d~R^{+t_{-}(x_0)-\alpha_{-} \ln \frac{x_1}{x_0}-1}~~~~~{\rm for}~~R < 1~,
    \label{Rline2}
\end{eqnarray}
where
\begin{eqnarray}
    \alpha_{+} \sim \alpha_{-} &\sim& 0~~~~~~~~~{\rm for}~~x_1 > x_0~,
    \label{alphaH}\\
    \alpha_{+} \sim \alpha_{-} &\neq & 0~~~~~~~~~{\rm for}~~x_{{\rm min}} < x_1 < x_0~,
    \label{alphaM}\\
    t_{+}(x_0) - t_{-}(x_0) &=& \mu~.
    \label{mu}
\end{eqnarray}
By combining extended Gibrat's law (\ref{Rline1}) -- (\ref{mu}) with 
the detailed balance (\ref{Detailed balance}),
we have kinematically derived the following 
distribution function in the high and middle scale region uniformly
\begin{eqnarray}
    P_1(x_1) = C~ {x_1}^{-\left(\mu+1\right)}~
    e^{-\alpha \ln^2 \frac{x_1}{x_0}}~~~~~~~~~{\rm for}~~x_1 > x_{{\rm min}}~,
    \label{HandM}
\end{eqnarray}
where $\alpha = (\alpha_{+} + \alpha_{-})/2$.

All these findings are enormously important
for the progress of econophysics.
Above derivations are, however, valid only in the economic equilibrium
where the detailed balance (\ref{Detailed balance}) holds.
As the next step, the dynamics should be established.
For this aim, we must investigate long-term
economic data in which dynamical transitions are observed.
Unfortunately, it is difficult to obtain personal income or company size data
for a long period. 

In Ref.~\cite{Kaizoji}, Kaizoji has reported
that the distribution of Japanese land prices
has similar features with one of personal income and company size.
The database of the assessed value of land is
available on the Ministry of Land, Infrastructure and Transport Government
of Japanese World-Wide Web site \cite{Web}.

In this paper, 
we analyze the high scale region data on the assessed value of land
in a 23-year period (1983 -- 2005)
in order to investigate the dynamical behavior.
As a result,
we obtain a dynamical equation which non-equilibrium economic systems should satisfy.

\section{Data of land prices and the law of detailed quasi-balance}
\label{sec-Data of land prices and the law of detailed quasi-balance}
\indspace
In Japan, land is a very important asset
which is distinguished from building.
The transition of land prices has a possibility
to influence the economy.
In fact, Japanese economy experienced bubble term (1986 -- 1991)
caused by the abnormal rise of land prices.

\begin{figure}[hbp]
 \centerline{\epsfxsize=0.8\textwidth\epsfbox{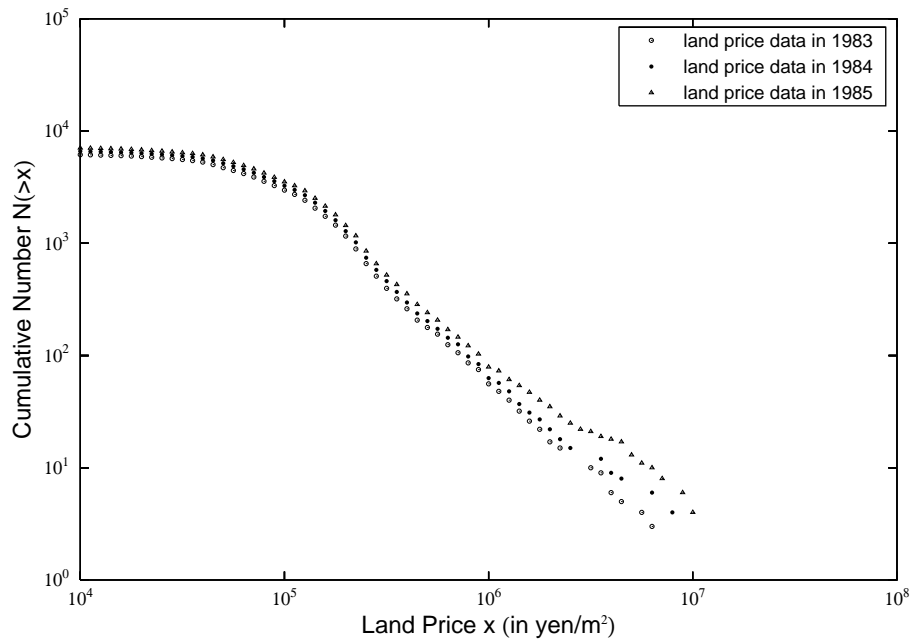}}
 \caption{Cumulative number distribution $N(>x)$ of land prices
 in 1983 -- 1985.
 The number of the data points is ``6,262'', ``6,706`` and ``7,169'', respectively.}
 \label{PreBubbleDistribution}
\end{figure}

\begin{figure}[hbp]
 \centerline{\epsfxsize=0.8\textwidth\epsfbox{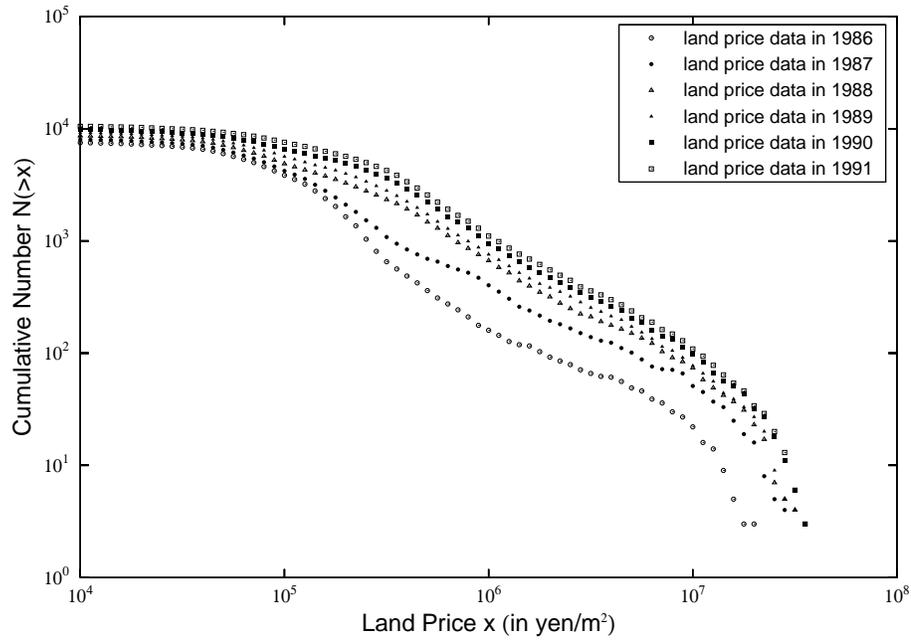}}
 \caption{Cumulative number distribution $N(>x)$ of land prices
 in 1986 -- 1991.
 The number of the data points is ``7,613``, ``8,120'', 8,739'',
``9,321'', ``9,927'' and ``10,603'', respectively.}
 \label{BubbleDistribution}
\end{figure}

\begin{figure}[hbp]
 \centerline{\epsfxsize=0.8\textwidth\epsfbox{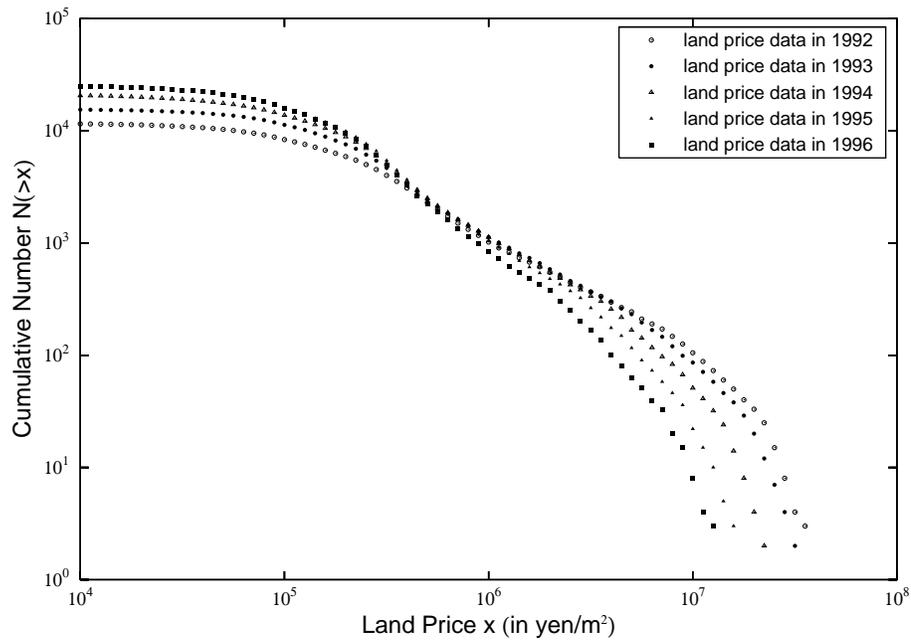}}
 \caption{Cumulative number distribution $N(>x)$ of land prices
 in 1992 -- 1996.
 The number of the data points is ``11,641'', ``15,538'',
``20,807'', ``24,753'' and ``25,091'', respectively.}
 \label{AfterBubbleDistribution}
\end{figure}

\begin{figure}[hbp]
 \centerline{\epsfxsize=0.8\textwidth\epsfbox{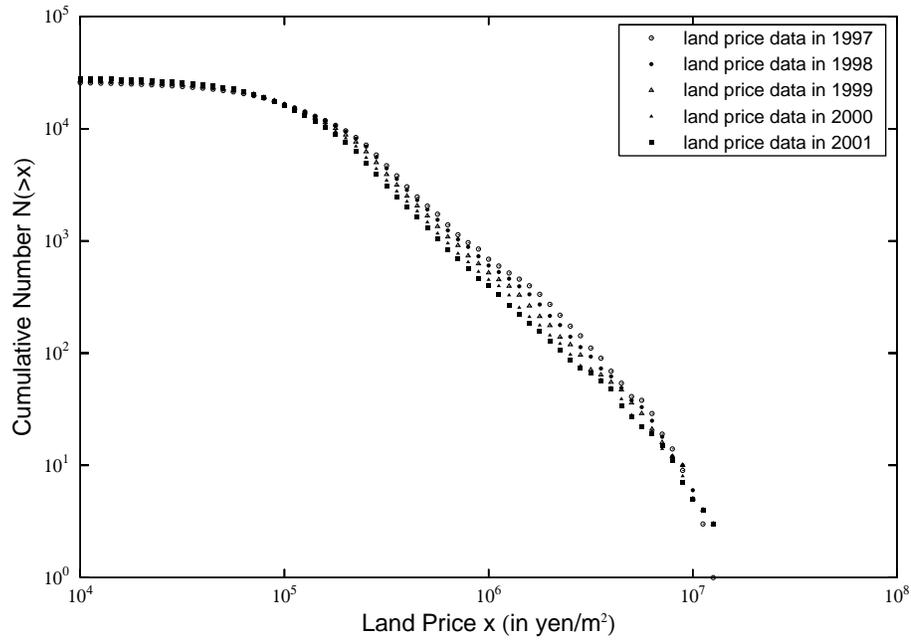}}
 \caption{Cumulative number distribution $N(>x)$ of land prices
 in 1997 -- 2001.
 The number of the data points is ``25,845'', ``26,524'',
``27,073'', ``27,738'' and ``28,294'',  respectively.}
 \label{StaticDistribution}
\end{figure}

\begin{figure}[hbp]
 \centerline{\epsfxsize=0.8\textwidth\epsfbox{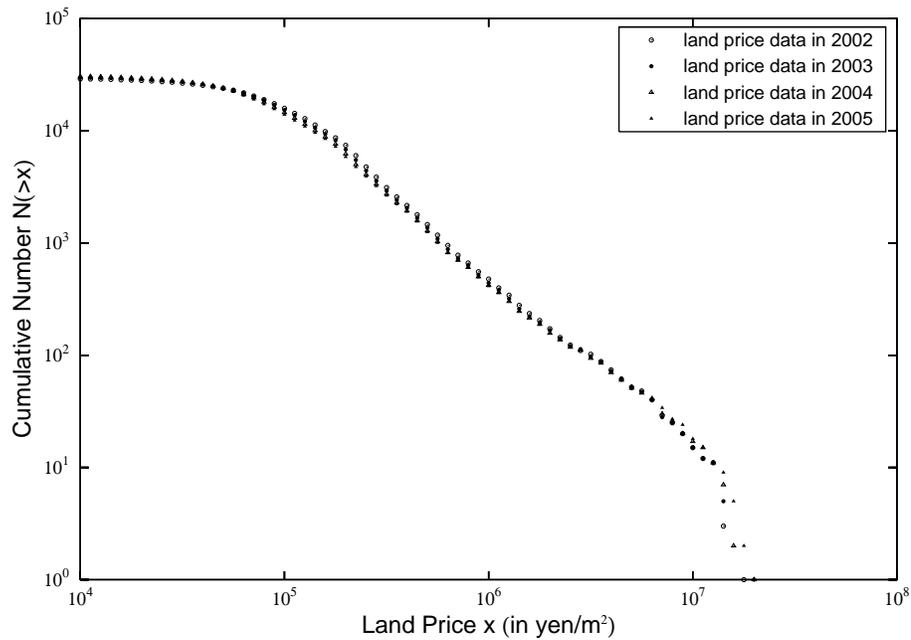}}
 \caption{Cumulative number distribution $N(>x)$ of land prices
 in 2002 -- 2005.
 The number of the data points is  ``29,279'', ``30,138'',
``30,775'' and ``31,230'', respectively.}
 \label{quasiStaticDistribution}
\end{figure}

The assessed value of land indicates 
the standard land prices evaluated by Ministry of Land, Infrastructure, 
and Transport.
The investigation is undertaken on each piece of land assessed once a year.
We employ the database of the assessed value of land
covering the 23-year period from 1983 to 2005.
The distributions of land prices are shown in
Fig.~\ref{PreBubbleDistribution}, \ref{BubbleDistribution},
\ref{AfterBubbleDistribution}, \ref{StaticDistribution} 
and \ref{quasiStaticDistribution}.
The number of data points of land prices increases gradually,
because the database 
only contains data points which exist in the 2005 evaluation.

\begin{figure}[htb]
 \centerline{\epsfxsize=0.8\textwidth\epsfbox{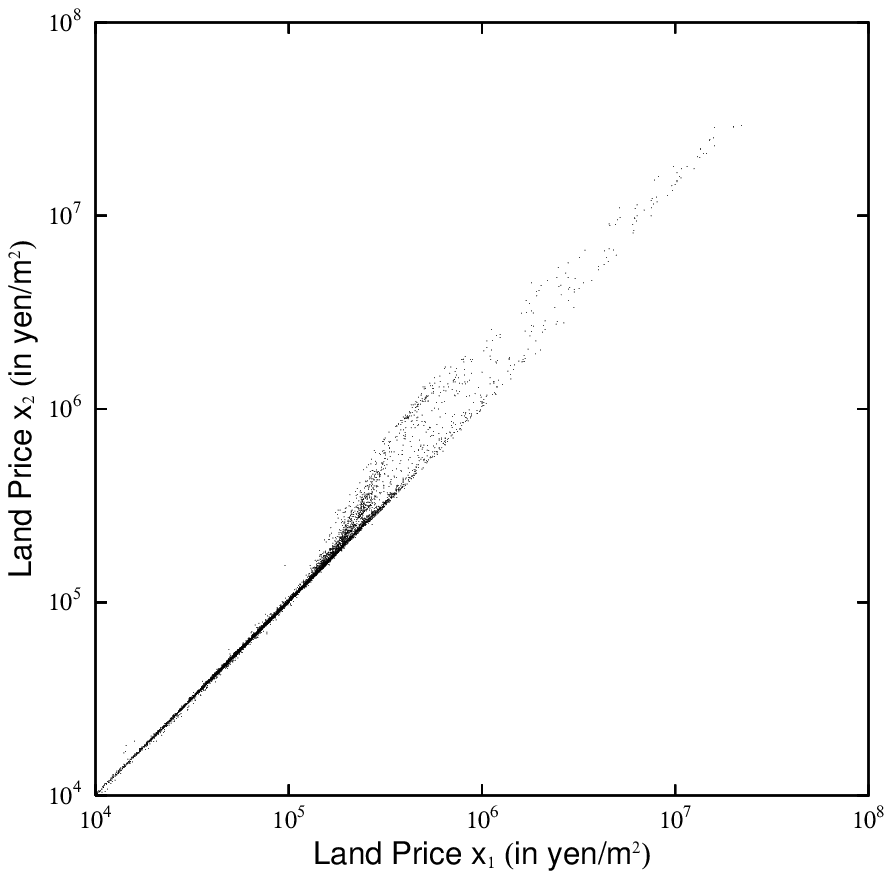}}
 \caption{The scatter plot of all pieces of land assessed in the database,
 the values of which in 1986 ($x_1$) and 1987 ($x_2$) exceeded $10^4 ~{\rm yen}/m^2$.
 The number of data points is ``1,603''.}
 \label{86vs87}
\end{figure}

\begin{figure}[htb]
 \centerline{\epsfxsize=0.8\textwidth\epsfbox{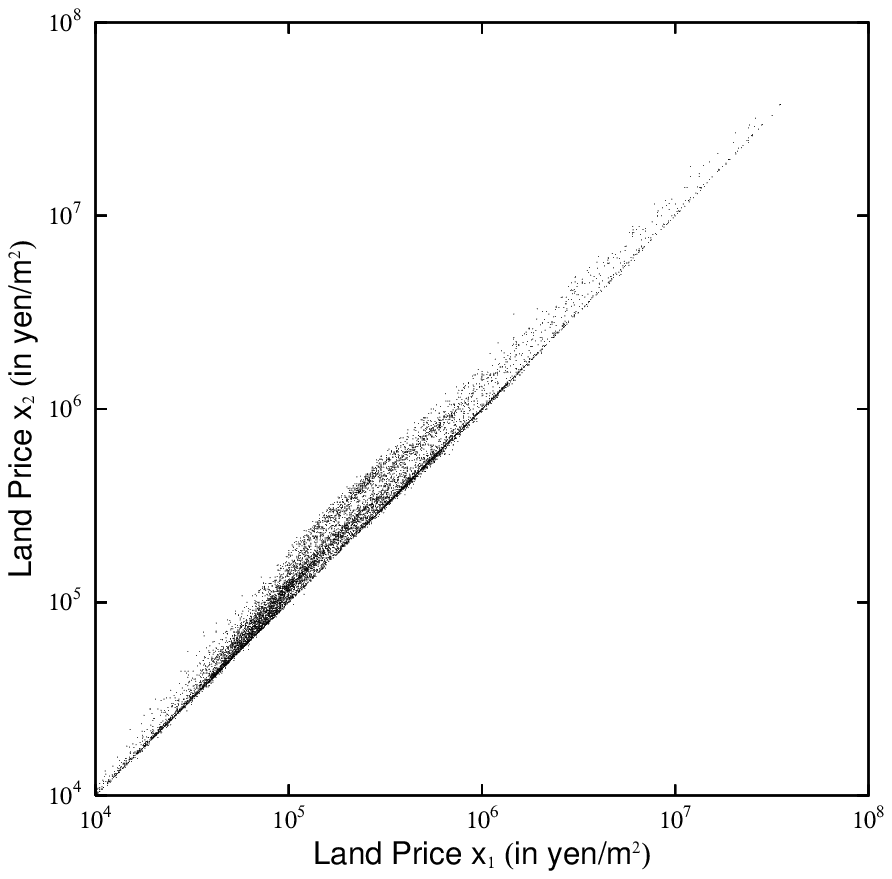}}
 \caption{The scatter plot of all pieces of land assessed in the database,
 the values of which in 1989 ($x_1$) and 1990 ($x_2$) exceeded $10^4 ~{\rm yen}/m^2$.
 The number of data points is ``3,886''.}
 \label{89vs90}
\end{figure}

\begin{figure}[htb]
 \centerline{\epsfxsize=0.8\textwidth\epsfbox{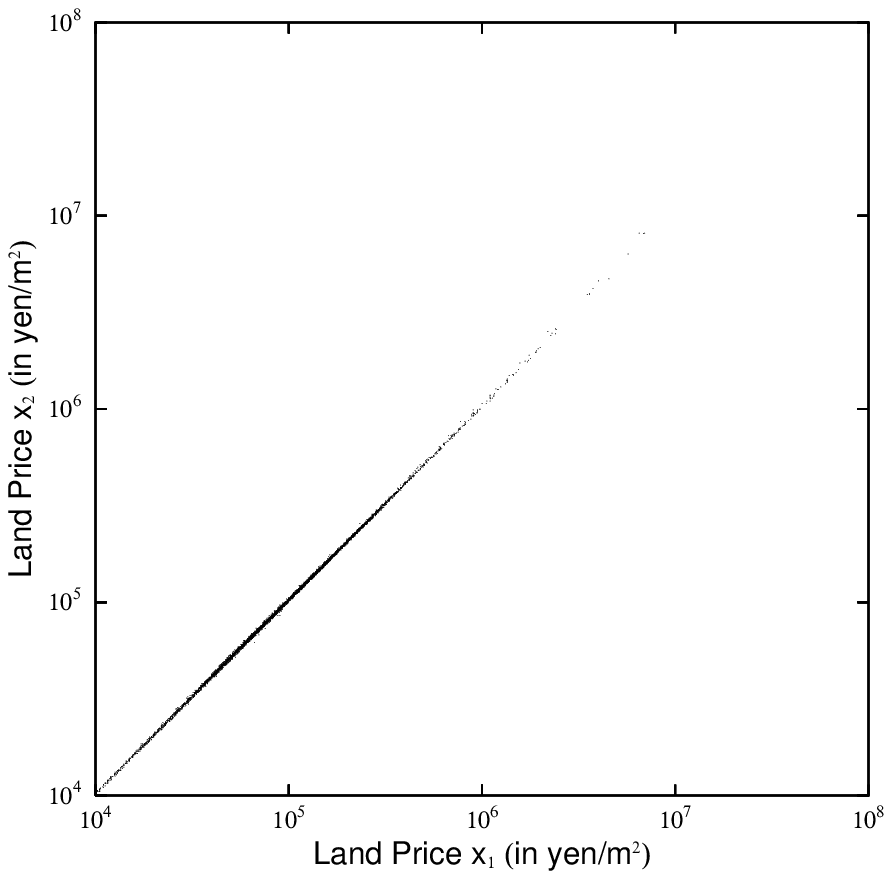}}
 \caption{The scatter plot of all pieces of land assessed in the database,
 the values of which in 1983 ($x_1$) and 1984 ($x_2$) exceeded $10^4 ~{\rm yen}/m^2$.
 The number of data points is ``1,160''.}
 \label{83vs84}
\end{figure}

\begin{figure}[htb]
 \centerline{\epsfxsize=0.8\textwidth\epsfbox{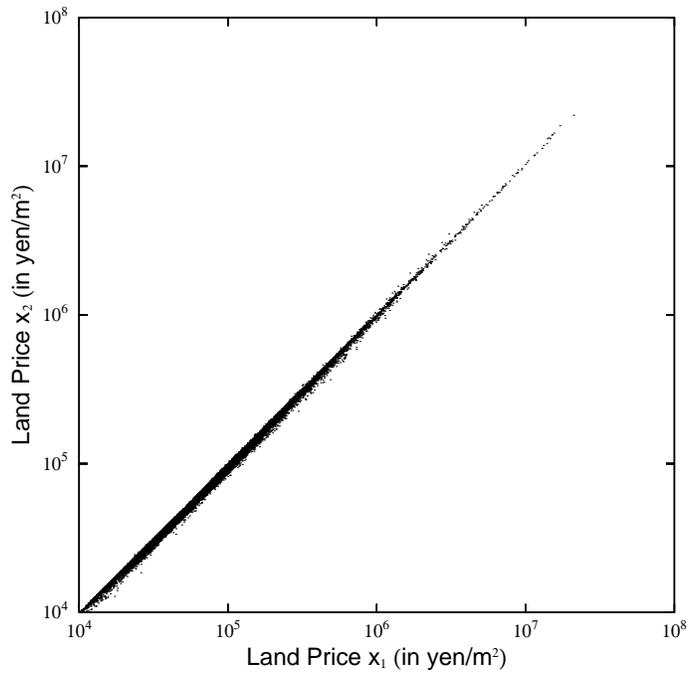}}
 \caption{The scatter plot of all pieces of land assessed in the database,
 the values of which in 2004 ($x_1$) and 2005 ($x_2$) exceeded $10^4 ~{\rm yen}/m^2$.
 The number of data points is ``5,739''.}
 \label{04vs05}
\end{figure}

From Fig.~\ref{PreBubbleDistribution} -- \ref{quasiStaticDistribution},
the power-law is confirmed in the high scale region.
For each year, we estimate Pareto index $\mu$ in the range of land prices from
$2 \times 10^5$ to $10^7~{\rm yen}/m^2$
where Pareto law holds approximately.
Annual change of Pareto index $\mu$ from 1983 to 2005 is represented
in Fig.~\ref{VaryingParetoIndex}.
In this period, Pareto index has changed annually.
This means that the system is not in equilibrium
and the detailed balance (\ref{Detailed balance}) does not hold.
Actually, the breakdown is observed in the scatter plot
of all pieces of land assessed in the database
(Fig.~\ref{86vs87} and \ref{89vs90} for instance).
There is no $x_1 \leftrightarrow x_2$ symmetry in 
Fig.~\ref{86vs87} and \ref{89vs90} obviously.
On the other hand,
the detailed balance ($x_1 \leftrightarrow x_2$ symmetry)
is observed approximately in Fig.~\ref{83vs84} and \ref{04vs05}
for instance.

\begin{figure}[htb]
 \centerline{\epsfxsize=0.8\textwidth\epsfbox{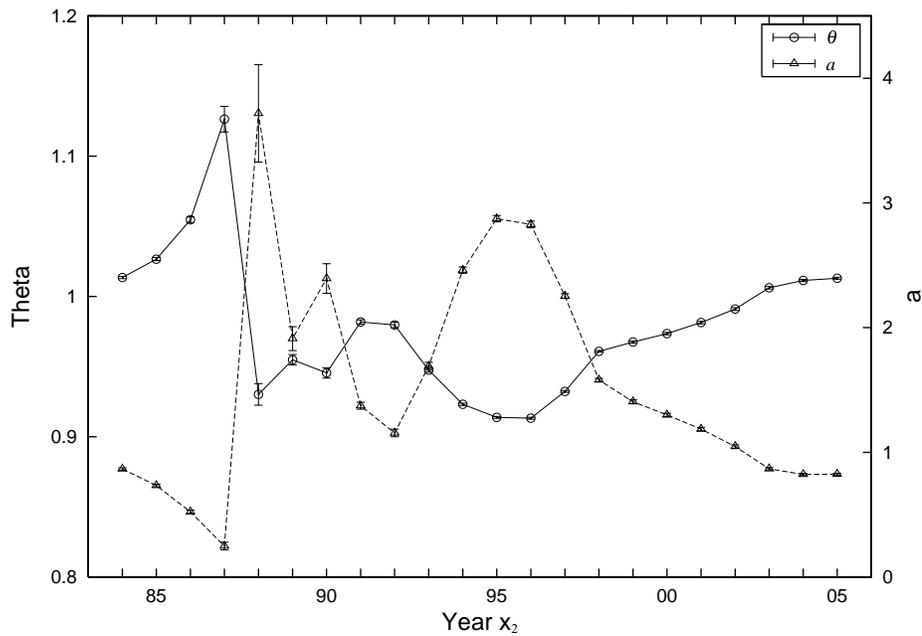}}
 \caption{Annual change of $\theta$ and $a$ of Eq.~(\ref{DqB}) in
 the year $(x_1, x_2) = (1983, 1984) -- (2004, 2005)$.
 }
 \label{DqB}
\end{figure}

Here we make a simple assumption that
the symmetry of the joint pdf $P_{1 2}(x_1, x_2)$ is represented
as a regression line fitted by least-square method as follows
\begin{eqnarray}
    \log_{10} x_2 = \theta~\log_{10} x_1 + \log_{10} a~.
    \label{Line}
\end{eqnarray}
The detailed balance (\ref{Detailed balance}) has the special symmetry, 
$\theta = a = 1$.
For each scatter plot, we measure $\theta$, $a$
and the result is shown in Fig.~\ref{DqB}.

From this symmetry ($a {x_1}^{\theta} \leftrightarrow x_2$), 
we extend the detailed balance (\ref{Detailed balance}) to
\begin{eqnarray}
    P_{1 2}(x_1, x_2) 
    = P_{1 2}( \left( \frac{x_2}{a} \right)^{1/{\theta}}, a~{x_1}^{\theta})~.
    \label{Detailed quasi-balance}
\end{eqnarray}
In this paper, we call this law the detailed quasi-balance.

\section{Dynamically varying Pareto index}
\label{sec-Dynamically varying Pareto index}
\indspace
In this section,
we derive the relation between the change of Pareto index $\mu$ and 
$\theta$ in Eq.~(\ref{Line}) by
extending the detailed balance (\ref{Detailed balance}) to
the detailed quasi-balance (\ref{Detailed quasi-balance}).
We assume Gibrat's law in the high scale region,
because the number of data points is insufficient to observe Gibrat's law.

Due to the relation of
$P_{1 2}(x_1, x_2)dx_1 dx_2 = P_{1 R}(x_1, R)dx_1 dR$
under the change of variables from $(x_1, x_2)$ to $(x_1, R)$,
these two joint pdfs are related to each other,
\begin{eqnarray}
    P_{1 R}(x_1, R) = {x_1}^{\theta} P_{1 2}(x_1, x_2)~,
\end{eqnarray}
where we use a modified ratio $R \equiv x_2/{x_1}^{\theta}$.
From this relation, the detailed quasi-balance (\ref{Detailed quasi-balance})
is rewritten in terms of $P_{1 R}(x_1, R)$ as follows:
\begin{eqnarray}
    P_{1 R}(x_1, R) 
    = a R^{-1} P_{1 R}(\left( \frac{x_2}{a} \right)^{1/{\theta}}, a^2 R^{-1})~.
\end{eqnarray}
Substituting the joint pdf $P_{1 R}(x_1, R)$ for the conditional probability $Q(R|x_1)$
defined in Eq.~(\ref{conditional}),
the detailed quasi-balance is expressed as
\begin{eqnarray}
    \frac{P_1(x_1)}{P_1(\left( {x_2}/{a} \right)^{1/{\theta}})} 
    &=& \frac{a}{R} 
    \frac{Q(a^2 R^{-1}|\left( {x_2}/{a} \right)^{1/{\theta}})}{Q(R|x_1)}
    \label{DqB and Gibrat0}\\
    &=& \frac{a}{R} 
    \frac{Q(a^2 R^{-1})}{Q(R)} \equiv G(a)~.
    \label{DqB and Gibrat}
\end{eqnarray}
We have used Gibrat's law in Eq.~(\ref{DqB and Gibrat0}).

By expanding Eq.~(\ref{DqB and Gibrat}) around $R=a$, the
following differential equation is obtained 
\begin{eqnarray}
    a ~G'(a) ~\theta~ P_1(x_1) 
        + x_1~ P'(x_1) = 0~.
\end{eqnarray}
The solution is given by
\begin{eqnarray}
    P_1(x_1) = C_1 ~{x_1}^{-a ~G'(a) ~\theta}~.
    \label{HandM}
\end{eqnarray}
Here we consider two power-law distributions
$N_1(> x_1) \propto {x_1}^{\mu_1}$ and $N_2(> x_2) \propto {x_2}^{\mu_2}$,
which lead
\begin{eqnarray}
    P_1(x_1) &=& C_1 ~{x_1}^{-\mu_1-1}~,\\
    \label{Pareto0}
    P_2(x_2) &=& C_2 ~{x_2}^{-\mu_2-1}~.
    \label{Pareto2}    
\end{eqnarray}
From Eq.~(\ref{HandM}) and (\ref{Pareto0}), we identify that
$a ~G'(a) ~\theta = \mu_1+1$.
On the other hand,
by taking
$x_1 \longrightarrow \left( x_2/a \right)^{1/{\theta}}$ transformation,
Eq.~(\ref{HandM}) is rewritten as
\begin{eqnarray}
    P_1(\left( \frac{x_2}{a} \right)^{1/{\theta}}) 
    = \frac{C_1}{a^{1/\theta}} ~{x_2}^{-a ~G'(a)}~.
    \label{Pareto2-2}  
\end{eqnarray}
Identifying that
$C_2 = C_1/a^{1/\theta}$ and
$P_1(\left( x_2/a \right)^{1/{\theta}}) = P_2 (x_2)$, 
we obtain $a ~G'(a) = \mu_2+1$.
Consequently the relation between $\mu_1$, $\mu_2$ and $\theta$ is expressed as
\begin{eqnarray}
    \frac{\mu_1+1}{\mu_2+1} = \theta~.
    \label{Ratio2}  
\end{eqnarray}

This is the dynamical equation which non-equilibrium economic systems satisfy.
We confirm that the empirical data satisfy this correlation in Fig.~\ref{Ratio}.

\section{Conclusion}
\label{sec-Conclusion}
\indspace
In this paper,
we have investigated the dynamical behavior
of non-equilibrium system in the high scale region
by employing data on the assessed value of land
in 1983 -- 2005 Japan.
By assuming the detailed quasi-balance (\ref{Detailed quasi-balance})
and Gibrat's law (\ref{Gibrat}), we have derived
a relation between the change of Pareto index and a symmetry in
the detailed quasi-balance.
The relation (\ref{Ratio2}) has been confirmed in the empirical data nicely.
The result in this paper is considered to be applied to other long-term economic data.
The equation (\ref{Ratio2}) is the dynamical one which non-equilibrium systems satisfy.

\begin{figure}[hbp]
 \centerline{\epsfxsize=0.8\textwidth\epsfbox{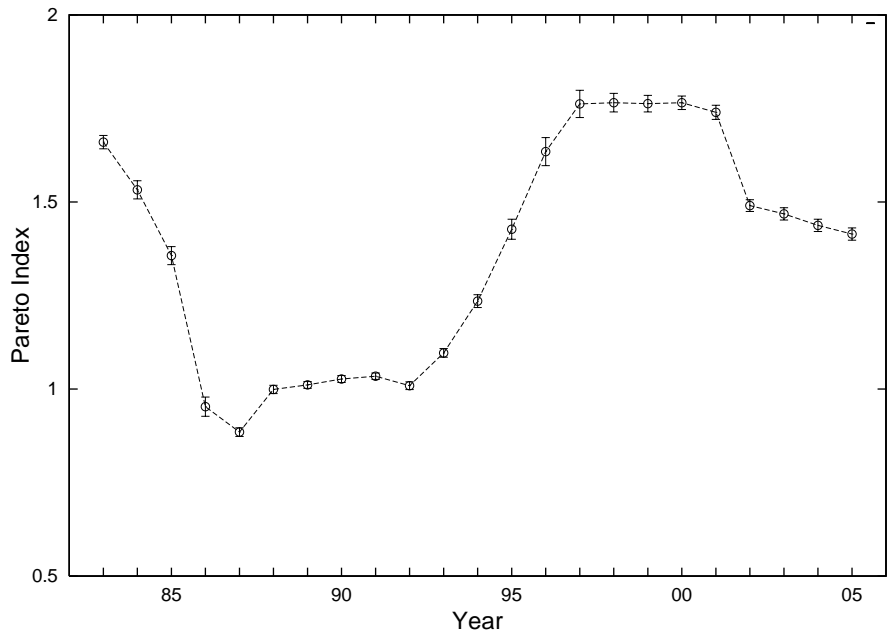}}
 \caption{Annual change of Pareto index $\mu$  from 1983 to 2005.
 }
 \label{VaryingParetoIndex}
\end{figure}

\begin{figure}[htb]
 \centerline{\epsfxsize=0.8\textwidth\epsfbox{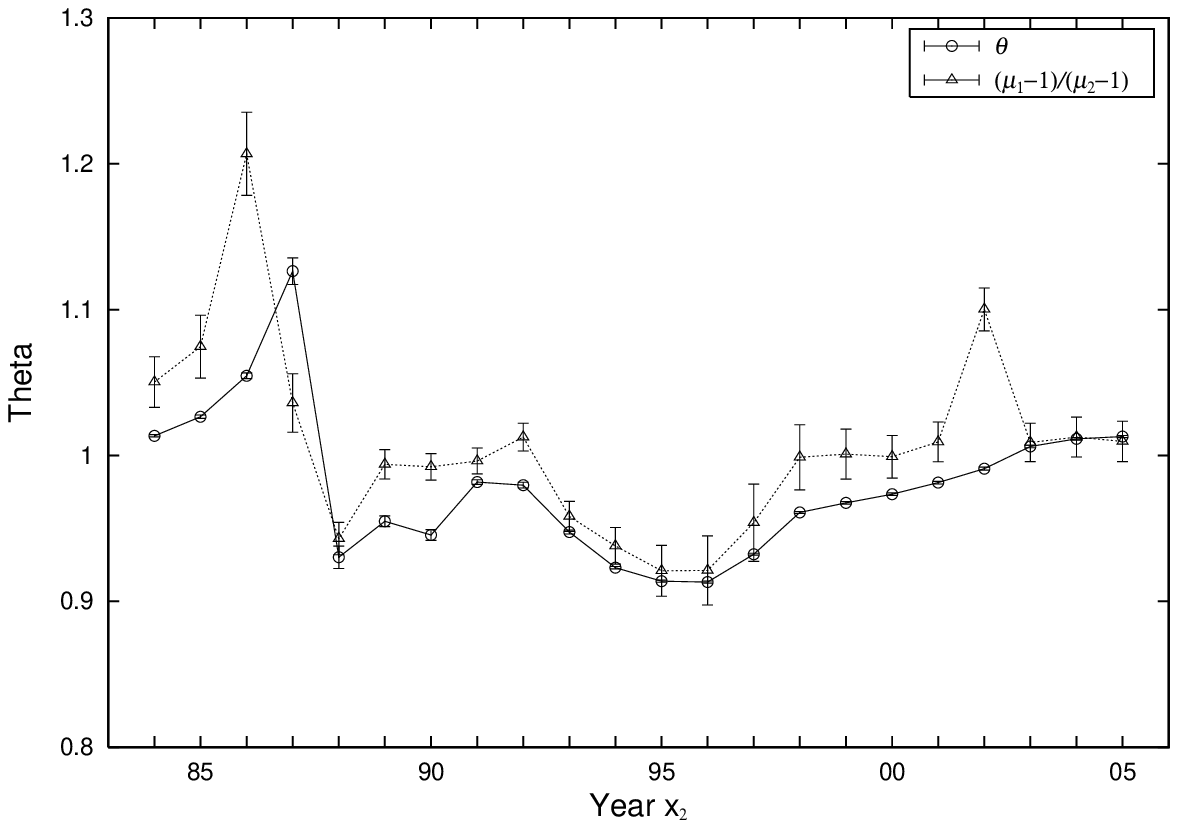}}
 \caption{Annual change of $\theta$ and $(\mu_1+1)/(\mu_2+1)$
 in the year $(x_1, x_2) = (1983, 1984) -- (2004, 2005)$.
}
 \label{Ratio}
\end{figure}

We should comment on two separations between $\theta$ and $(\mu_1 + 1)/(\mu_2 + 1)$ 
in Fig.~\ref{Ratio}.
An abrupt jump of Pareto index between 1985 and 1986 (2001 and 2002)
is observed in Fig.~\ref{VaryingParetoIndex}.
This means that the system changes vigorously
in this period, where
the symmetry is not represented as Eq.~(\ref{Line}).
Nevertheless, the dynamical equation (\ref{Ratio2}) is
valid in almost all the other quasistatic periods.

The finding in this paper
is a first step to
explain varying macro phenomena
by using laws in non-equilibrium systems.
For the next step,
we should investigate the dynamical behavior in the middle scale region.

%





\begin{thebibliography}{99}
\bibitem{MS}
    R.N.~Mategna and H.E.~Stanley, An Introduction to Econophysics, 
    Cambridge University Press, UK, 2000.
\bibitem{Pareto}
    V.~Pareto, Cours d'Economique Politique, Macmillan, London, 1897.
\bibitem{ASNOTT}
    H.~Aoyama, W.~Souma, Y.~Nagahara, H.P.~Okazaki, H.~Takayasu and M.~Takayasu, 
    Fractals 8 (2000) 293;\\
    W.~Souma, Fractals 9 (2001) 463.
\bibitem{Yakovenko}
    A.~Dr$\check{a}$gulescu and V.M.~Yakovenko, Physica A299 (2001) 213.
\bibitem{FSAKA}
    Y.~Fujiwara, W.~Souma, H.~Aoyama, T.~Kaizoji and M.~Aoki,
    Physica A321 (2003) 598;\\
    H.~Aoyama, W.~Souma and Y.~Fujiwara, Physica A324 (2003) 352;\\
    Y.~Fujiwara, C.D.~Guilmi, H.~Aoyama, M.~Gallegati and W.~Souma,
    Physica A335 (2004) 197;\\
    Y.~Fujiwara, H.~Aoyama, C.D.~Guilmi, W.~Souma and M.~Gallegati,
    Physica A344 (2004) 112;\\
    H.~Aoyama, Y.~Fujiwara and W.~Souma, Physica A344 (2004) 117.
\bibitem{Gibrat}
    R.~Gibrat, Les inegalites economiques, Paris, Sirey, 1932. 
\bibitem{Badger}
    W.W.~Badger, in: B.J.~West (Ed.), Mathematical Models as a Tool for the Social Science, 
    Gordon and Breach, New York, 1980, p. 87;\\
    E.W.~Montrll and M.F.~Shlesinger, J. Stat. Phys. 32 (1983) 209.

\bibitem{Stanley1}
    M.H.R.~Stanley, L.A.N.~Amaral, S.V.~Buldyrev, S.~Havlin, H.~Leschhorn, P.~Maass, M.A.~Salinger and H.E.~Stanley,
    Nature 379 (1996) 804;\\
    L.A.N.~Amaral, S.V.~Buldyrev, S.~Havlin, H.~Leschhorn, P.~Maass,
    M.A.~Salinger, H.E.~Stanley and M.H.R.~Stanley, J. Phys. (France) I7 (1997) 621;\\
    S.V.~Buldyrev, L.A.N.~Amaral, S.~Havlin, H.~Leschhorn, P.~Maass,
    M.A.~Salinger,  H.E.~Stanley and M.H.R.~Stanley, J. Phys. (France) I7 (1997) 635;\\
    L.A.N.~Amaral, S.V.~Buldyrev, S.~Havlin, M.A.~Salinger and H.E.~Stanley, 
    Phys. Rev. Lett. 80 (1998) 1385;\\
    Y.~Lee, L.A.N.~Amaral, D.~Canning, M.~Meyer and H.E.~Stanley,
    Phys. Rev. Lett. 81 (1998) 3275;\\
    D.~Canning, L.A.N.~Amaral, Y.~Lee, M.~Meyer and H.E.~Stanley,
    Economics Lett. 60 (1998) 335. 
\bibitem{TTOMS}
    H.~Takayasu, M.~Takayasu, M.P.~Okazaki, K.~Marumo and T.~Shimizu, cond-mat/0008057,
    in: M.M.~Novak (Ed.), Paradigms of Complexity, World Scientific, 2000, p. 243.
\bibitem{Aoyama}
    H.~Aoyama, 9th Annual Workshop on Economic Heterogeneous Interacting Agents (WEHIA 2004);\\
    H.~Aoyama, Y.~Fujiwara and W.~Souma, The Physical Society of Japan 2004 Autumn Meeting.
\bibitem{Ishikawa3}
    A.~Ishikawa, Derivation of the distribution from extended Gibrat's law, physics/0508178. \bibitem{TSR}
    TOKYO SHOKO RESEARCH, LTD., http://www.tsr-net.co.jp/.
\bibitem{Kaizoji}
    T.~Kaizoji, Physica A326 (2003) 256.
\bibitem{Web}
    The Ministry of Land, Infrastructure and Transport Government of Japan's World-Wide Web site,
    http://nlftp.mlit.go.jp/ksj/.
\end{thebibliography}
\end{document}